\documentclass[a4paper,11pt]{article}
\title{\textbf{Double differential cross sections for
ionization of hydrogen atoms by electron impact in hyperspherical
partial wave theory }}
\date{}
\usepackage{amssymb}
\usepackage{latexsym}
\usepackage{graphicx}
\begin{document}
\setcounter{section}{0}
\renewcommand{\thesection}{\Roman{section}.}
\setcounter{equation}{0}
\renewcommand{\theequation}{\arabic{equation}}
\maketitle
\begin{center}
J. N. Das and S. Paul\\
{\it{Department of Applied Mathematics, University College of
Science, 92, Acharya Prafulla Chandra Road, Calcutta - 700 009,
India}}\\
\end{center}

\begin{abstract}

    The calculated double differential cross sections for 60eV
energy, where the cross sections are also maximum, agree with the
measure results of Shyn (1992) in a much better way compaired to
other theories for such energies.\\\\

\end{abstract}

e-mail : jndas@cucc.ernet.in \\\\

\pagebreak
\noindent

    With the availability of of results of some recent
calculations of Bray (2002) and R\"oder et al (2003) and Das
\textit{et al} (2003 a,b) the status of theoretical studies has
reached a new epoch. Even then much remains to be done. For
example, although the convergent close coupling (CCC) approach
(Bray, 2000) gives beautifully the total, single, and the triple
differential cross sections (for some kinematic conditions), it
miserably fails in reproducing the double differential cross
sections (DDCS) for intermediate and high energies. As regards the
DDCS results the same is true for 3C theory (Berakdar and
Klar,1993). Results of external complex scaling (ECS) approach ,
now a famous approach, are not available, possibly due to
difficulties. In this letter we present DDCS results of a
calculation in hyperspherical partial wave approach (HPW) for 60eV
energy, a typical intermediate energy, for which the cross
sections are known to have maximum values.

      The hyperspherical partial wave approach has been described
in detail in (Das \textit{et al},2003a) and also in greater detail
in (Das,1998). In this approach the scattering state wave function
is expanded in symmetrized hyperspherical harmonics (Das,1998; Das
\textit{et al},2003a) resulting in sets of infinite coupled
equations in infinite number of radial wave functions in hyper
radius R. After truncating these equations to some finite
$N_{mx}^L$ numbers keeping the same number of radial wave
functions one has to solve these over the semi-infinite interval
(0, $\infty$) . In this calculation we have chosen $N_{mx}^L$ =40
for L=0 and 1 and $N_{mx}^L$ =80 for other L values. Here L values
up to 12 has been found to be practically sufficient. For
convenience in solution of the equations over (0, $\infty$ ) we
divide it into three subintervals (0, $\Delta$ ), ($\Delta,
R_\infty$ ), ($R_\infty, \infty$ ). As usual we have chosen
$\Delta$ , 5.0 a.u.. The asymptotic range parameter $R_\infty$ has
to be properly chosen. We have set our codes in such a way that
this parameter $R_\infty$ may be made to take values thousands of
atomic units. This is possible only for another approach, the
hyperspherical R-matrix method with semi classical outgoing waves
(HRM-SOW) approach of Malegat \textit{et al}(2000). Now inclusion
of a large number of L values, at the same time makes the
calculation very time consuming. So with our moderate resources we
have included values of L up to 12 and chosen $R_\infty$ about 300
a.u.. With this choice we have moderately converged results.

    In our present calculation our main interest is in the DDCS
results. However we have also calculated single and total
ionization cross sections to have a complete picture. The single
differential cross sections have approximately converged except at
the two ends, where energy of one of the electrons is below 2eV
and where contamination with high Rydberg states gives
unacceptable very high cross section results as in ECS calculation
by Flux method (Baertschy \textit{et al} 2001).

    In figure 1 we have presented the calculated SDCS results,
fourth order parabolic fitted cross section curves (singlet,
triplet and total) have also been included. From these fitted SDCS
results total cross sections (singlet, triplet and their sum) have
also been calculated, as also the asymmetry parameter A. These
results have been presented in table 1. Finally in figure 2 we
have presented the DDCS results which is our main objective, here
DDCS results have been normalized in a way that the integrated
single differential cross sections agree with the parabolic fitted
values. The results beautifully agree with the measured results
for energies below 6eV of the ejected electron. Particularly the
agreement with experiment for 3eV and 4eV are remarkable (see
fig.2). Here the normalization factors are also very close to
unity. No earlier calculation could produce such accurate results
(see for example figures 8 and 12 of Bray (2000) and figures 5-7
of Berakdar and Klar (1993) for some other intermediate values.
Unfortunately their results for this particular energy of 60eV are
not available.

    In conclusion it may be said that the hyperspherical partial
wave theory is successful in reproducing the DDCS results,
generally, in a far better manner compared to other theories. If
we recall its capability in describing ionization events from
close to threshold (viz. 1eV, 0.5eV and 0.3eV excess energies as
in Das \textit{et al} 2003b) to a few Rydberg energy (Das 2002;
Das \textit{et al} 2003a) and the present one then the
hyperspherical partial wave theory may be claimed to have an edge
over other theories. With the availability of better computational
facilities this claimed may be better established. More
experimental results, both double and triple differential cross
sections, for intermediate energies are urgently needed for better
assessment of different sophisticated theories.\\

\noindent
\textbf{Acknowledgments}
\noindent

    One of the authors ( S.Paul ) is grateful
to CSIR for providing a research fellowship.\\

\pagebreak

\noindent

\underline{\bf{References}}
\begin{flushleft}

[1] Baertschy M, Rescigno T N, Isaacs W A, Li X and  McCurdy C W
    2001 Phys. Rev. A \textbf{63} 022712.\\

[2] Berakdar J and Klar H 1993 J. Phys. B \textbf{26} 3891.\\

[3] Bray I, 2002 Phys. Rev. Lett. \textbf{89} 273201.\\

[4] Bray I, 2000 Aust. J. Phys. \textbf{53} 355.\\

[5] Das J N 1998 Pramana- J. Phys. \textbf{50} 53.\\

[6] Das J N 2002 J. Phys. B  \textbf{35} 1165.\\

[7] Das J N, Paul S and Chakrabarti K 2003a Phys. Rev. A
\textbf{67} 042717.\\

[8] Das J N, Paul S and Chakrabarti K 2003b J. Phys. B
(communicated).\\

[9] Fite WL and Brackmann RT 1958 Phys. Rev. \textbf{112} 1141.\\

[10] Fletcher GD, Alguard MJ, Gray TJ, Wainwright PF, Lubell MS,
Raith W and Hughes VW 1985 Phys. Rev. A \textbf{31} 2854.\\

[11] Malegat L, Selles P and Kazansky A K 2000 Phys. Rev.
     Lett.{\bf{21}} 4450.\\

[12] R\"oder J, Baertschy M and Bray I 2003 Phys. Rev. A
\textbf{67} 010702.\\

[13] Shah MB, Elliot DS and Gilbody HB 1987 J. Phys. B \textbf{20}
3501.\\

[14] Shyn TW 1992 Phys. Rev. A \textbf{45} 2951.\\

\end{flushleft}

\pagebreak

\noindent

\underline{\bf{Figure Captions}}\\

\noindent
     \textbf{Figure 1.} Single differential cross sections of
the present calculation as compared with the experimental
results of Shyn(1992).\\
\noindent
     \textbf{Figure 2.} Double differential cross sections of the
present calculation as compared with the experimental results of
Shyn (1992). The results have been normalized by factors,
indicated in the figures, for the integrated single differential
cross section to agree with the parabolic fitted results of figure 1.\\

\pagebreak
\begin{center}
\underline{{\bf{Table}}}
\end{center}

\noindent \textbf{Table 1.} Singlet, triplet and the total
ionization cross section (in 10$^-8$ cm) together with the spin
asymmetry parameter A of the present calculation at 60 eV energy
as compared with the experimental results. Starred results
correspond to the energy indicated in the bracket.\\

\begin{tabular}{lllll}
\hline
 & $\sigma_0$ & $\sigma_1$ & $\sigma_T$ & $A$ \\
Present & 35.7 & 53.5 & 89.2 & 0.200 \\
Experiment\\
Shyn: & - & - & 87.0 & -\\
Fite and Brackmann: & - & - & 72.0 & - \\
Shah \textit{et al}: & - & - & 51.3 & - \\
Fletcher \textit{et al}: & - & - & - & 0.236$^* \pm 0.021$ (57
eV)\\ \hline
\end{tabular}
\pagebreak
\begin{figure}[h]
\begin{center}
\includegraphics[width=14cm]{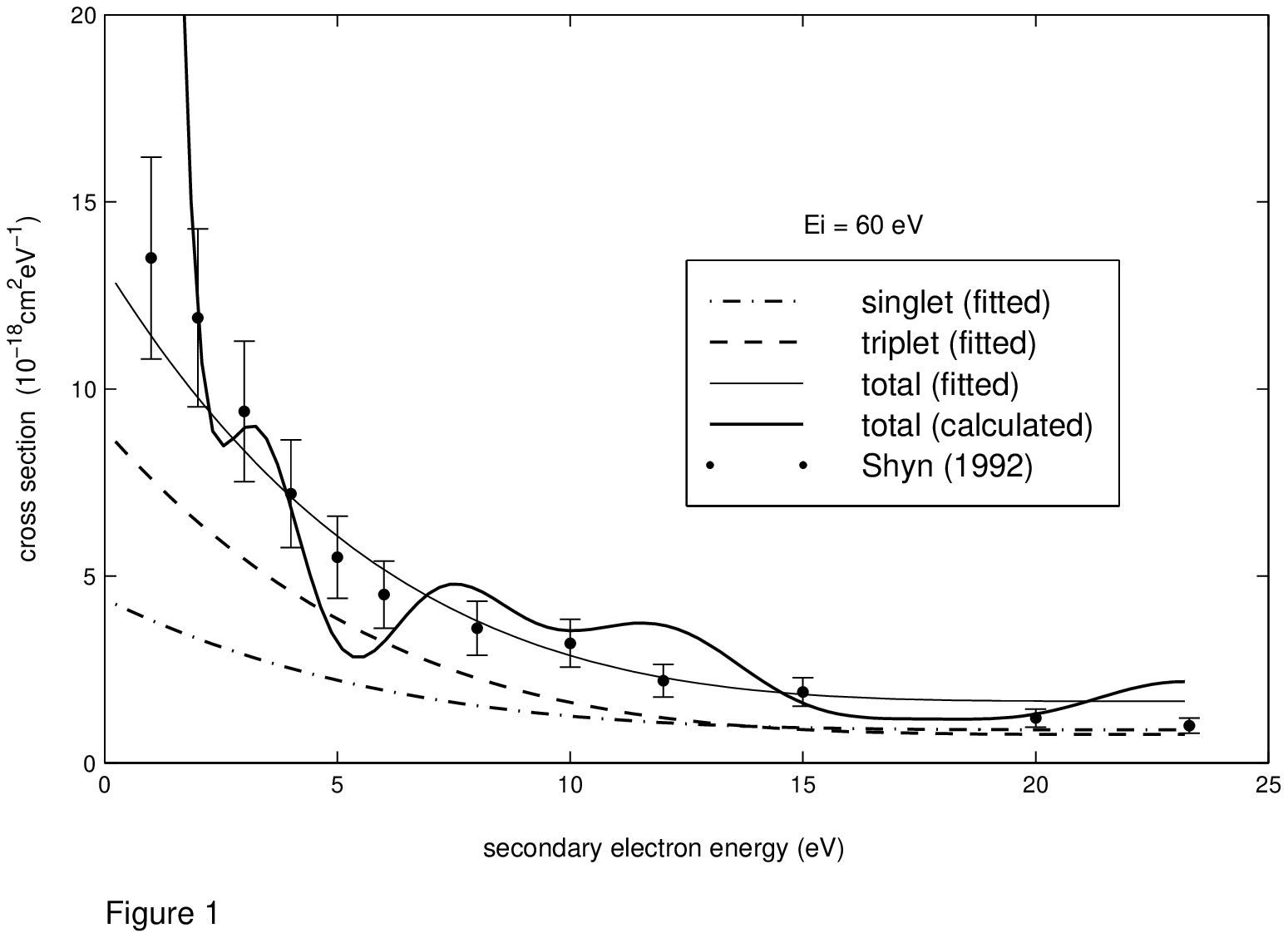}
\end{center}
\end{figure}

\begin{figure}[h]
\begin{center}
\includegraphics[width=14cm]{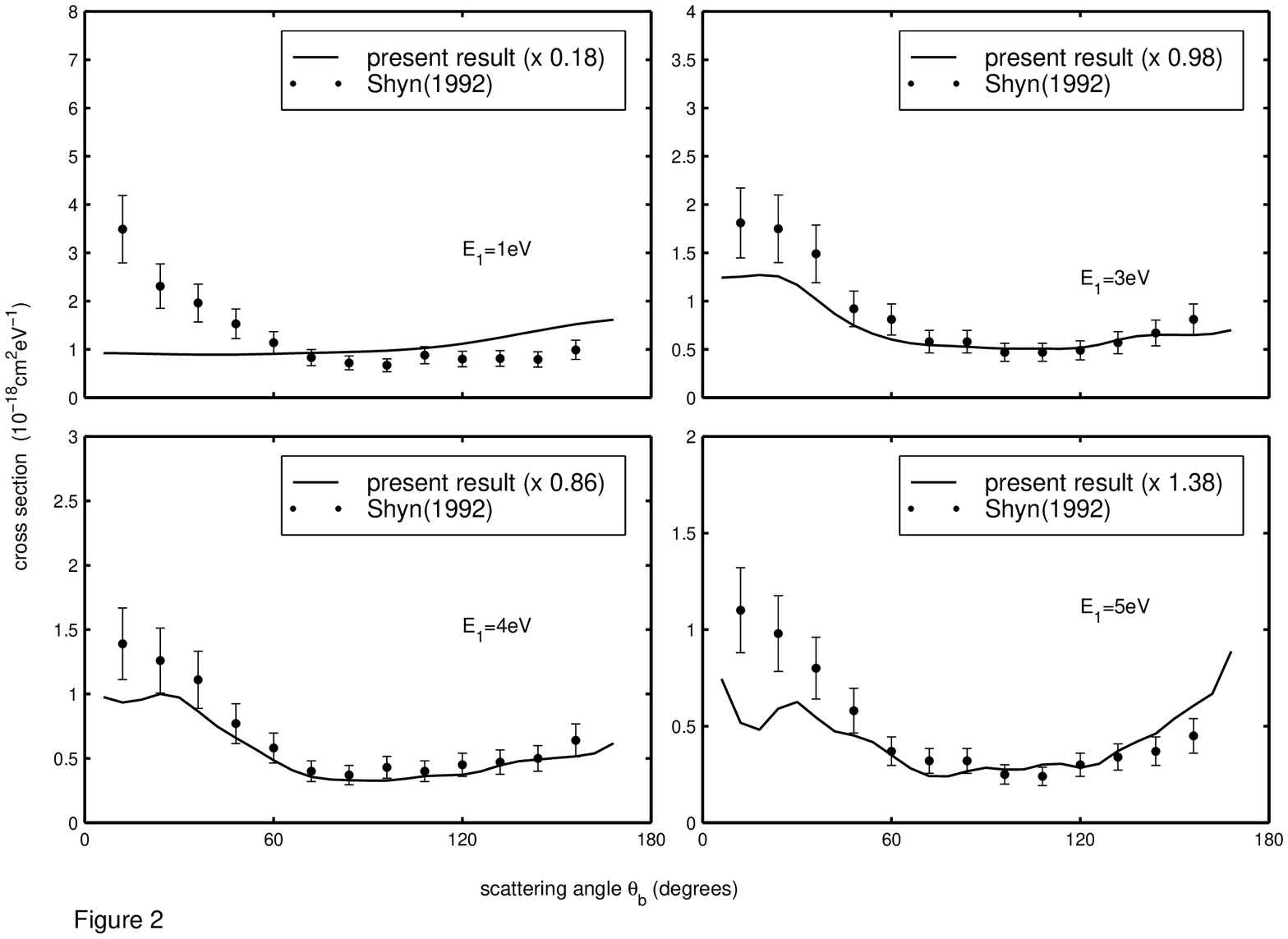}
\end{center}
\end{figure}
\end{document}